Manuscript of the published article:

# Study of nitrogen ion doping of titanium dioxide films

Raul Ramos, Diego Scoca, Rafael Borges Merlo, Francisco Chagas Marques, Fernando Alvarez, Luiz Fernando Zagonel*

"Gleb Wataghin" Institute of Physics

University of Campinas – UNICAMP

13083-859, Campinas, São Paulo, Brazil

This study reports on the properties of nitrogen doped titanium dioxide ($TiO_2$) thin films considering the application as transparent conducting oxide (TCO). Sets of thin films were prepared by sputtering a titanium target under oxygen atmosphere on a quartz substrate at 400 or 500°C. Films were then doped at the same temperature by 150 eV nitrogen ions. The films were prepared in Anatase phase which was maintained after doping. Up to 30at% nitrogen concentration was obtained at the surface, as determined by *in situ* x-ray photoelectron spectroscopy (XPS). Such high nitrogen concentration at the surface lead to nitrogen diffusion into the bulk which reached about 25 nm. Hall measurements indicate that average carrier density reached over $10^{19}$ $cm^{-3}$ with mobility in the range of 0.1 to 1 $cm^2V^{-1}s^{-1}$. Resistivity about $3.10^{-1}$ $\Omega cm$ could be obtained with 85% light transmission at 550 nm. These results indicate that low energy implantation is an effective technique for $TiO_2$ doping that allows an accurate control of the doping process independently from the $TiO_2$ preparation. Moreover, this doping route seems promising to attain high doping levels without significantly affecting the film structure. Such approach could be relevant for preparation of $N:TiO_2$ transparent conduction electrodes (TCE).

## Graphical abstract

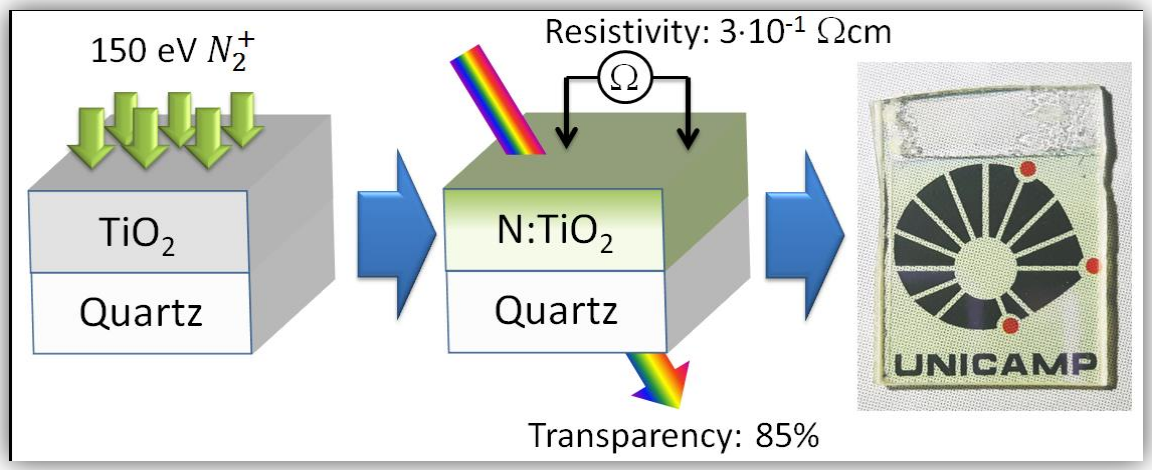

## Highlights

- A two-step process for preparation of N:TiO$_2$ transparent conductor is proposed.
- Low energy nitrogen ions are used after Anatase thin film deposition.
- Approach allows excellent control of crystal, optical and electronic properties.
- Resistivity as low as $3 \cdot 10^{-1}$ Ωcm while transparency at 550nm is about 85%.
- High temperatures enable thermal diffusion of Nitrogen inside Anatase film.

# 1 Introduction

The increasing demand of energy efficiency and cost-effectiveness for display and energy technologies pushes continuously towards the search of new materials to surpass industry standards. In flat panel displays, light emitting devices and some solar cells, efficiency is linked to the performance of the transparent conducting electrodes (TCEs) used which allows front electrical contacts and simultaneously letting visible light in or out of the device. While tin-doped indium oxide (ITO) is an industry standard TCE, it presents a high cost linked to indium scarcity. Most alternatives available today, such as ZnO or $SnO_2$, has interesting niche applications. New and better TCEs, both in terms of cost and efficiency, are of great interest for several wide or niche applications.

Since titanium oxide ($TiO_2$) doped with niobium has been proposed as TCE by Hasegawa group see refs. [1,2], several studies discussed the optical and electrical properties of $TiO_2$ doped with Nb, Ta, W, and N [3,4,5,6]. Moreover, it has been shown that, by doping $TiO_2$ with nitrogen, it is possible to reduce its optical gap and favor catalytic activity with visible light, with considerable interest for water splitting, among other applications [7,8]. Since then, several studies explored the properties of nitrogen doped titanium oxides (mainly Anatase and Rutile) prepared in various ways with respect to its optical, catalytic and transport properties.

Nitrogen doped $TiO_2$ has already been synthetized by reactive sputtering and by post treatments with ammonia or ion implantations, among others. Using electron cyclotron resonance plasma sputtering under $O_2$ and $N_2$ gases, H. Akazawa showed that it is possible to continuously control carrier concentration and obtained films with a resistivity of 0.2 $\Omega$cm with a maximum transparence in the visible of about 80%, but the films were frequently amorphous, while large crystalline grains might favor better conductivity at similar transparency [4,6]. Using reactive d.c. magnetrons sputtering in an Ar+$O_2$+$N_2$ gas mixture, N. Martin et al. obtained about 25$\Omega$cm with about 30% transmission, depositing $TiN_xO_y$. Again, in this study, crystal structure was difficult to control and, besides Rutile and Anatase, even $Ti_3O_5$ was observed [9]. In another work by J.-M. Chappé et al., also using d.c. reactive magnetron sputtering, prepared $TiO_xN_y$ films with visible light transmittance ranging from very low to nearly 80%, with a resistivity ranging from $10^{-3}$ $\Omega$cm to 50 $\Omega$cm and with a complex crystal structure where Anatase was not the majority phase present [10]. Given the inherent difficulty in controlling independently composition and crystal structure/quality in reactive sputtering, splitting the processes in two parts is an interesting alternative. In this approach, Anatase or Rutile samples can be prepared and doped *a posteriori* with nitrogen by, for instance, ion implantation or $NH_3$ gas [11,12,13]. Using this approach, H. Shen *et al*. showed that implantation with 200 eV nitrogen ions successfully doped Anatase nanoparticles and enhanced photocatalytic efficiency without changing the crystal structure [13].

Considering that ion doping of $TiO_2$ could be relevant for TCE preparation, in this work, we deposited Anatase thin films at 400 and 500°C and then doped by low energy nitrogen ion implantation at 150eV from a simple laboratory ion gun. Such process allows doping pure Anatase thin films with controllable nitrogen amounts and following closely its properties. By heating the sample during the ion implantation, we could allow the diffusion of nitrogen from the surface into the bulk, thus developing a dopant profile. The results indicate that low energy nitrogen doping of Anatase is a promising route for preparation of this material, N:$TiO_2$ or $TiN_xO_y$ as a TCO.

## 2 Experimental Methods

Sample preparation was performed in two sequential steps. First, a Ti target was sputtered using an ion beam (Ion Beam Deposition) to grow a thin film on amorphous quartz substrate. Argon was used as inert gas for bombarding the Ti target at an energy of 1.5keV. During the deposition, a partial pressure of $2.5 \cdot 10^{-2}$ Pa of oxygen was maintained (chamber base pressure was about $2 \cdot 10^{-4}$ Pa). Such partial pressure results in Anatase films with well-defined x-ray diffraction peaks [14]. During the deposition, Argon partial pressure in the chamber was about $1 \cdot 10^{-2}$ Pa. During a second step, nitrogen ions were implanted at low energy, 150 eV, into the thin film surface. The ions were produced in a Kaufman cell fed with 5 sccm of Nitrogen and 0.5 sccm of hydrogen, resulting in $2.1 \cdot 10^{-2}$ Pa and $2.1 \cdot 10^{-3}$ Pa partial pressures in the chamber respectively (Argon and oxygen were not used in this step). Such sample preparation was performed in a custom-built system that features two ion guns (one pointing to a sputter target and the other to the sample holder) in one vacuum chamber that is directly connected to another chamber for *in situ* X-ray Photoemission Spectroscopy analysis. More details of the deposition system and its capabilities can be found in references [15] and [16]. Hydrogen was used in analogy with ref. [17] (see also references there in) to remove oxygen from the surface to make it more reactive for incoming nitrogen. Indeed, the formation enthalpy favors $TiO_2$ over TiN [18] and in principle residual oxygen gas and water vapor in the vacuum chamber could keep the surface partially oxidized preventing nitrogen intake. The ion gun points perpendicularly to the sample surface and is located about 30 cm from the sample. The samples were prepared at different substrate temperatures (for both steps): 400 and 500°C and with different implantation times: 0, 10, 30 and 60 minutes. Film thickness was evaluated by perfilometry and it ranges from 70 to 100 nm.

Just after preparation, samples were *in situ* analyzed in UHV by X-ray Photoemission Spectroscopy (XPS) using Al K$\alpha$ radiation. Spectra were fitted using Avantage software. Average inelastic mean free path for Anatase and kinetic energies from 900-1100 eV are estimated as about 2 nm [19]. X-ray diffraction (XRD) was performed using Cu K$\alpha$ and keeping incidence angle at 1°. In this geometry, the average penetration depth is estimated as 0.05 $\mu$m for Anatase [20]. Sheet resistance was measured using 4-probe technique. Mobility, resistivity and carrier concentration were determined by Hall measurements using the Van der Pauw method in an Ecopia-3000 device using a 0.55 T permanent magnet. For Hall measurements, indium was used to provide ohmic contacts. Optical transparency measurements were performed in an Agilent 8453 device which uses a CCD detector.

Resistance versus temperature measurement was performed for the sample deposited at 500°C and implanted for 60 minutes (at the same temperature). The measurement was carried out in a CTI Cryodine closed-cycle helium refrigerator in the temperature range from 80 K to 300 K. The electrical data was acquired using a Keithley model 2602A SourceMeter and the indium contacts previously used for Hall measurement, in the van der Pauw geometry. A constant current of 10 µA was applied between two contacts and the voltage was measured between the other two, in a parallel configuration. A complete thermal loop was carried out to confirm the reproducibility of the data.

Sample's morphology was studied by Atomic Force Microscopy (AFM) and Transmission Electron Microscopy (TEM). TEM analysis was performed in a JEOL 2100F TEM equipped with a Field Emission Gun (FEG) operating at 200 kV with an energy resolution of about 1 eV. EELS was obtained using a Gatan GIF Tridiem installed in this TEM and Gatan Digital Micrograph routines were used for quantification. The data were acquired in Scanning Transmission Electron Microscopy (STEM) mode in the form of spectrum lines (the electron beam is focused on the sample and a spectrum is acquired for each position along a line forming a bi-dimensional dataset). Topographic images of the sample's surface were taken with an Innova Bruker Atomic Force Microscope (AFM) in non-contact mode.

# 3    Results and Discussions

## 3.1    Composition and Structural characterization

The effectiveness of the ion implantation at 150 eV was demonstrated by the presence of large amounts of nitrogen at the surface, as observed by *in situ* XPS. Figure 1 shows XPS spectra for the sample prepared at 500°C for 60 minutes, with similar results for all other samples. Main features observed by XPS are expected for $TiO_2$ and for TiN. *In situ* XPS on $TiO_2$ samples grown at 400 and 500°C (not shown) are similar to those in ref [14] and [21], typical for Anatase film close to stoichiometry. It must be noted that absolute binding energies are not accurately known due to some degree of uncontrolled spectral shift that is attributed to sample charging. From the indicated decomposition into several proposed chemical bounds, it is possible to observe that nitrogen concentration is similar to that of oxygen and that a $TiO_xN_y$ alloy was created at the surface (TiN and $TiO_2$ components are observed in the Ti 2p spectrum). It is noteworthy in Figure 1(c) the presence of two XPS peaks for N 1s, one, smaller, at higher binding energy, and another, bigger, close to 396 eV (that is in turn composed of two peaks). Such smaller and bigger peaks are attributed to interstitial nitrogen and substitutional nitrogen, respectively [12,13,22,23]. In our case, interstitial N accounts to about 10% of the total amount of nitrogen observed on the surface, a much lower value when compared to ref. [8], which used $NH_3$ as nitrogen source, or ref. [13], which used 200 eV nitrogen ions without hydrogen. Therefore, depending on the incorporation route, different chemical locations are possible. This difference is relevant since depending on nitrogen site different diffusion mechanisms apply [12].

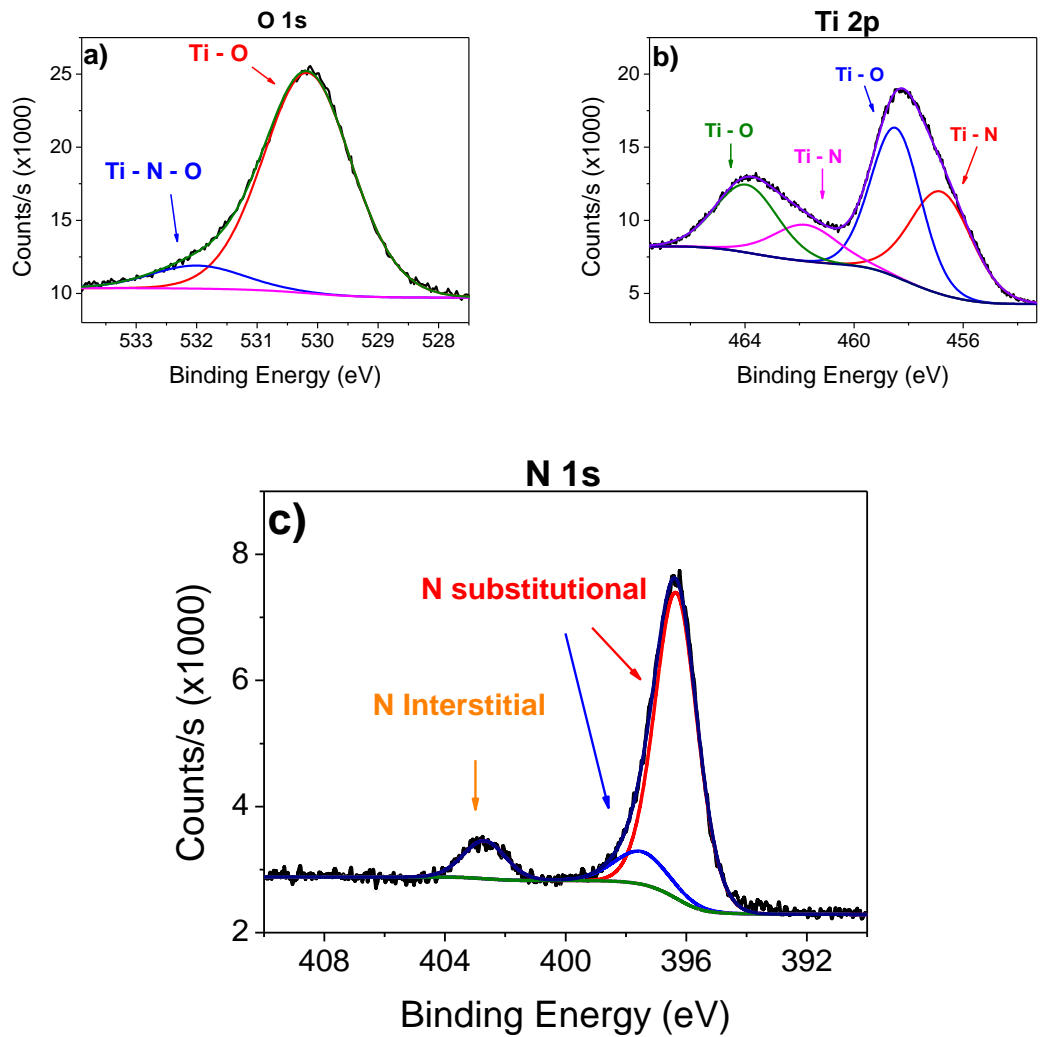

Figure 1: *In situ* x-ray photoemission spectra from sample grown at 500°C and implanted (at the same temperature) for 60 minutes with 150 eV nitrogen ions. The spectra include decomposition into components for different expected chemical bounds in the sample.

To evaluate if hydrogen was significantly affecting N 1s spectra, a sample was prepared without hydrogen gas during the implantation step. N1s peaks for samples prepared with and without hydrogen at 500°C and implanted for 60 minutes are shown in Figure 2. The spectra show that even without hydrogen the peak is still present and with similar (although smaller) ratio with respect to main N 1s peak. This shows that it does not depend on presence of hydrogen during the nitriding process, in contrast to literature suggestion [24]. However, as discussed below, total nitrogen concentration is smaller without hydrogen, indicating that hydrogen contributes to nitrogen incorporation at the surface, possibly by removing oxygen.[17]

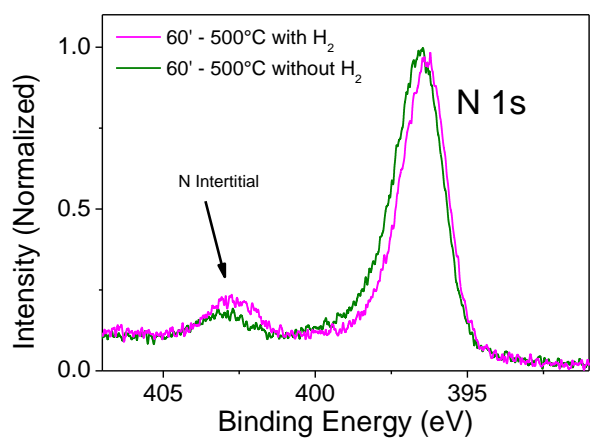

Figure 2: X-ray photoemission spectra are shown for samples prepared with and without hydrogen in the gas feed to the ion gun. The peak associated to Nitrogen in interstitial sites is observed in both samples.

To support the given interpretation of XPS results, SRIM simulations have been performed [25,26]. Simulations considered nitrogen ions (N+) on Anatase. Average penetration depth is about 0.9 nm while 90% of implanted nitrogen ion reaches a depth within 1.7 nm. These simulations indicate that XPS is probing exactly the implanted region and hence is suitable to investigate the nitrogen intake by the sample from the ion beam.

Following the procedures detailed in references [27] and [28], we used XPS results to calculate the elemental concentrations of surface components. The results are shows in Figure 3 for samples prepared at 400 and at 500°C with implantation times ranging from 0 to 60 minutes. It is observed that in the first few minutes of implantation, a significant nitrogen concentration builds up at the surface and after the concentration increases yet to reach about 33at.%. This is explained by the high reactivity of the nitrogen ion beam and the low diffusion coefficient of nitrogen into the interior of the thin film. In such scenario, we consider that a high nitrogen concentration builds up during the first moments and is maintained by the ion beam creating a high nitrogen chemical potential at the surface. This nitrogen concentration will be the driving force for nitrogen diffusion into the thin film. The extent of the diffusion will depend mainly on the temperature but also on several details of the thin film microstructure, such as vacancies, grain boundaries, stress, and so on. This process is in tight analogy to the plasma or ion beam nitriding of steels at low temperatures where nitrogen diffusion is also slow [29,30]. It is important to note that for both studied temperatures, with sufficient nitriding time, the surface builds a titanium oxynitride alloy with stoichiometry close to $TiO_1N_1$ (note such result applies only at the outer 2 nm of the thin film surface). It is also interesting that the sample prepared without hydrogen had a nitrogen concentration of only 25at.% while the sample prepared with hydrogen in the same conditions (500°C – 60 minutes) had 33at.% of nitrogen at the surface. Again this indicates that hydrogen may favor oxygen removal opening sites for nitrogen chemical adsorption and reaction, even if nitrogen arrives at 150eV at the surface and the sample is in high vacuum.

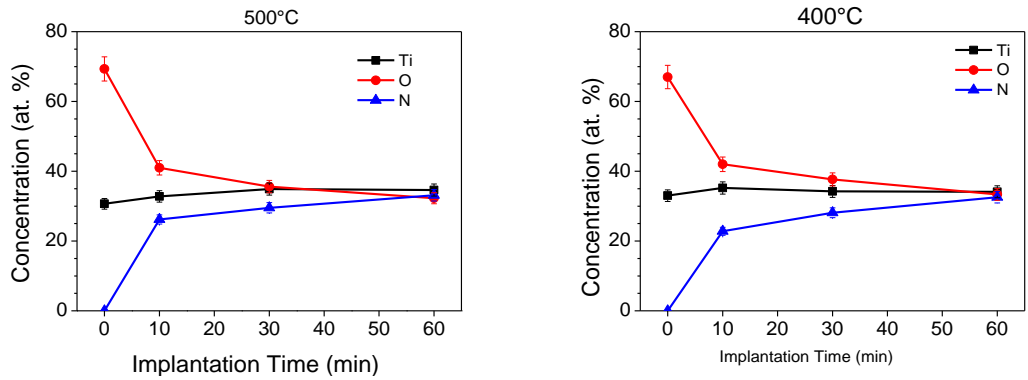

Figure 3: Nitrogen, Oxygen and Titanium concentration obtained by *in situ* XPS for samples prepared at 400° and 500°C for several implantation times. Lines are a guide to the eyes. Nitrogen concentration reaches about 33at.%.

X-ray diffractograms are shown in Figure 4 for samples prepared at 400 and 500° without nitrogen implantation and implanted for 10, 30 and 60 minutes, as before. It is observed that all samples display peaks associated with Anatase phase with considerable intensity indicating the crystalline nature of the thin films. Moreover, implantation with Nitrogen and Hydrogen does not disturb the crystal structure, similarly to what has been reported in the literature for 200eV nitrogen implantation into Anatase thin films [13]. The position of the Anatase (101) peak remains within (25.30±0.05)° for all diffractograms while reference Anatase (101) is expected at 25.33° according to ICSD 9852. It must be noted that samples are kept at deposition temperature during the implantation and are therefore annealed what could affect their crystalline structure. It is also noteworthy that peak ratios do not agree with expected values for Anatase powder and hence the films should have some texture [31].

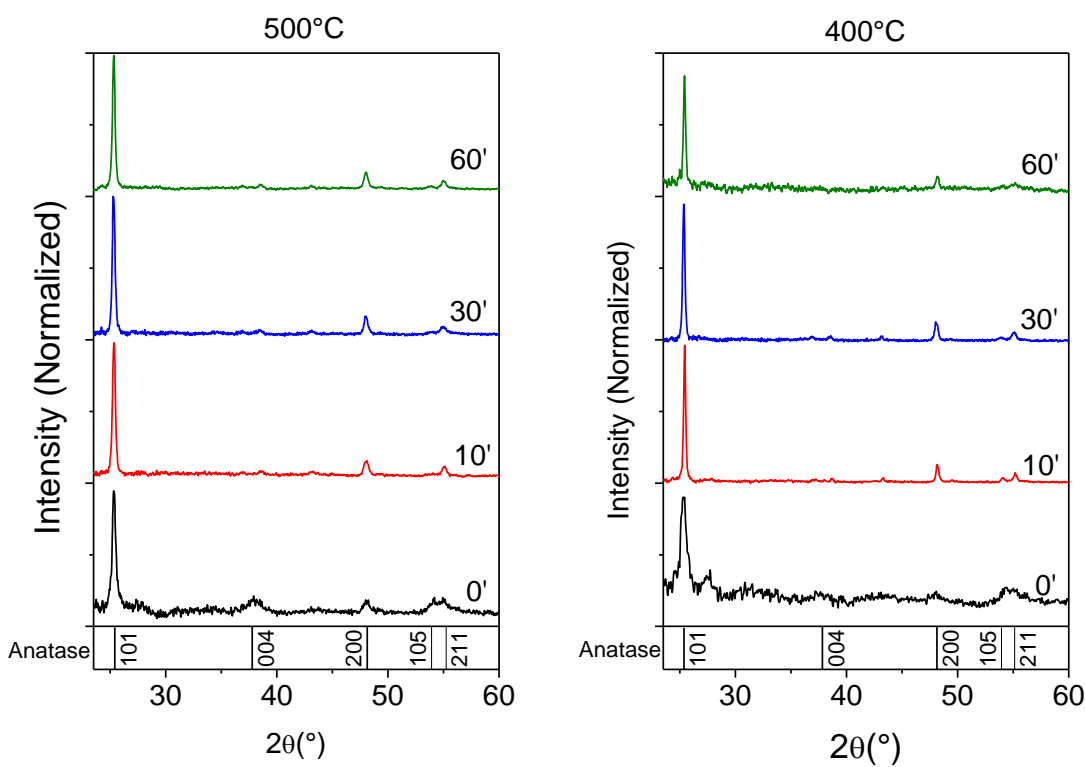

Figure 4: Diffractograms from grazing incidence X-ray Diffraction for $TiO_2$ films prepared at 400 and 500°C and implanted for the time indicated in each curve. Anatase lines are indicated in the bottom according to ICSD 9852.

Transmission Electron Microscopy was applied to determine the extent of nitrogen diffusion into the film from the surface. For that, we considered only the sample prepared at 500°C for 60 minutes, considering that other samples would have a shallower nitrogen penetration depth. Figure 5 shows a cross-section of the sample. Apart from the amorphous quartz substrate and the protective coating used for FIB lamella preparation, we can observe the N implanted $TiO_2$ thin film in two layers, on top a layer that apparently has been modified by the implantation/diffusion process and on the bottom the pristine Anatase film. HRTEM images indicate the presence of atomic planes and grains from bottom to top of the thin film, again confirming the Anatase film preserved its crystal structure even after nitrogen shallow implantation (shallower than 2nm from SRIM simulations) and subsequent diffusion.

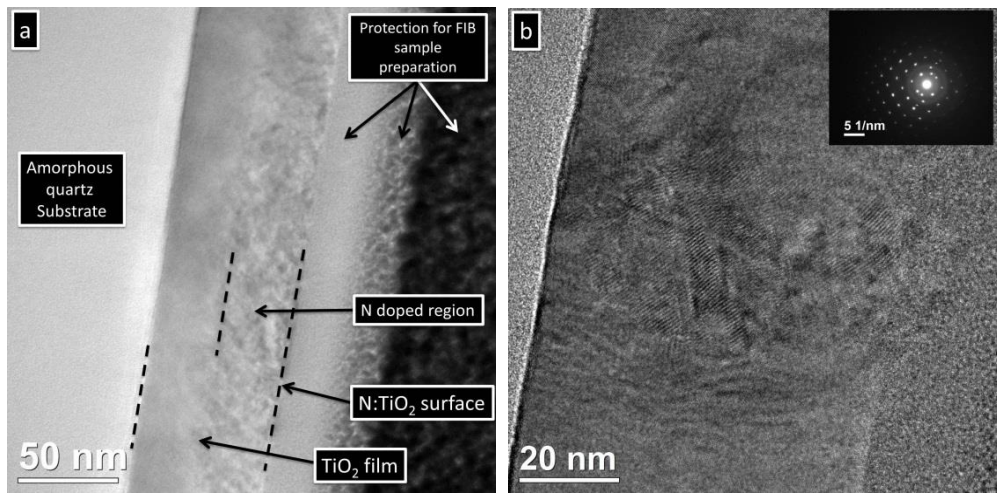

Figure 5: TEM cross-section micrograph from sample prepared at 500°C and implanted for 60 minutes. (a) A modified region at the surface is observed. (b) HR-TEM shows atomic planes from small grains from bottom to the top of the thin film. The insert shows a SAED pattern obtained nearby on a region with larger grains.

Electron energy loss spectroscopy was used to detect nitrogen and determine its profile in the sample cross-section. Figure 6(a) shows the profiles of nitrogen, oxygen and titanium from the surface to the interior of the thin film. The detection of nitrogen was difficult due to sample damage: apparently the electron beam removed nitrogen during beam exposure. For this reason (and also because of diffraction and thickness effects were not taken into account), the results in Figure 6(a) may underestimate the original concentration (due to damage even in reduced dose measurements) or other systematic error (due to the other mentioned effects). However, it can accurately be considered semi-quantitatively to measure the diffusion depth. In Figure 6(a), a complementary error function fitting is added to the nitrogen profile as a thin line. Despite the noise, it is clear that nitrogen is detected down to 25 nm or so (where estimated nitrogen concentrations decreases to 10% of its surface value). The presence of nitrogen is also clear in the fine-structure of Titanium and oxygen absorption edges, shown in Figure 6 (b). Again, a transition from one edge shape to the other is observed around 30 nm from the surface. Moreover, it is interesting to note that the obtained nitrogen profile didn't affect the crystal structure as observed by RH-TEM (Figure 5 (b)), that is, no amorphous layer was found despite the observed nitrogen concentration. Indeed, in some studies, the presence of nitrogen in reactive sputtering leads to amorphous N:TiO$_2$ films [4].

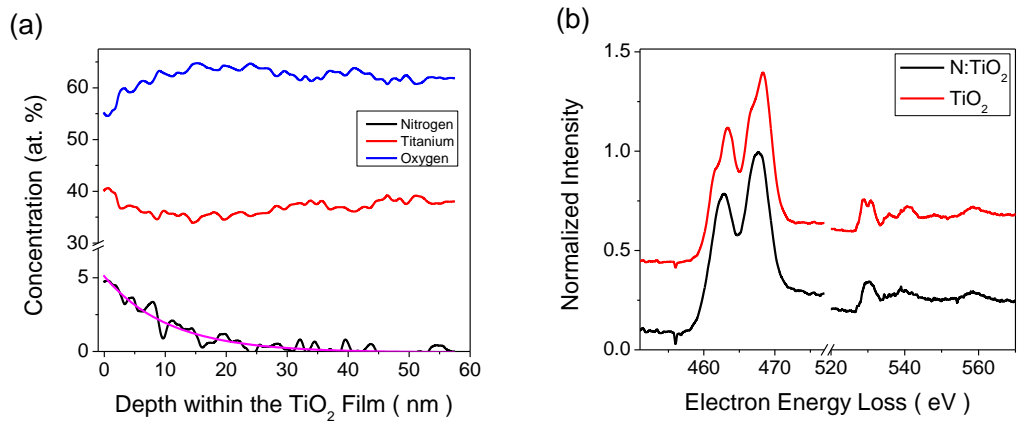

Figure 6: (a) Nitrogen, Oxygen and Titanium semi-quantitative profile determined by STEM-EELS on a cross-section lamella. A complementary error function fit was added to nitrogen profile. (b) Averaged EEL spectra of titanium 2p and oxygen 1s absorption edges indicating the difference from the upper (nitrogen doped) layer to the bottom (pristine) $TiO_2$.

Atomic force microscopy was used to gather a broader idea of the growth and also a clearer picture of the surface before and after ion implantation. The average surface roughness is about 1 nm for all samples, indicating a smooth growth of $TiO_2$ Anatase thin film by reactive sputtering and also that implantation at 150eV does not induce surface roughening. Illustrative results, for samples growth at 400°C without implantation and implanted for 30 minutes, are shown in Figure 7. Without implantation, the surface shows small grains having about 40-60 nm in diameter, but the height difference from peak to valley is just about 3 nm. After implantation, crystal grains are partially revealed, as indicated by arrows in Fig. 7(b), and their diameter is in fact about 200 to 300 nm, a result more consistent with TEM observations. As we consider that the ion implantation at 150eV or the annealing time didn't change the crystal structure, grains should have diameters in the hundreds of nanometers from the beginning of the deposition, but ion polishing was necessary to reveal the actual grains due to preferential sputtering of different crystal orientations [32].

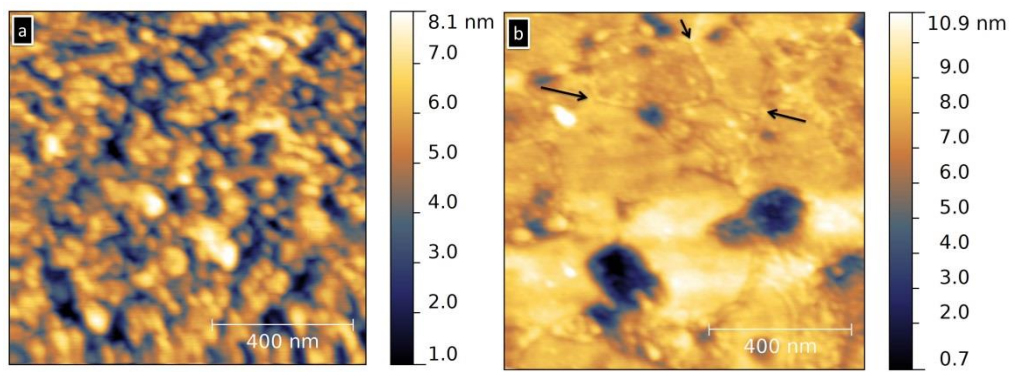

Figure 7: AFM images of the surface of samples implanted by 0 minutes (undoped) and 30 minutes, both prepared at 400°C, are shown is (a) and (b), respectively. Ion implantation reveals partially the grains by preferential sputtering. Arrows in (b) indicate grain boundaries.

From XRD, XPS, TEM and AFM results, it is possible to form the following picture of the implantation process: the ion beam is highly reactive and, with the help of hydrogen, creates a high surface chemical potential which is, together with the process temperature, the driving force for nitrogen diffusion into the thin $TiO_2$ film. Considering that a roughly constant nitrogen concentration builds up in the first minutes, a diffusion coefficient of nitrogen on anatase at 500°C can be calculated as approximately $2 \cdot 10^{-8} \mu m^2 s^{-1}$, considering the solution of Fick's second law for a constant surface concentration. R. G. Palgrave *et al.* fond similar values ($2.53 \cdot 10^{-8} \mu m^2 s^{-1}$) at 675° for rutile and also, analyzing very accurate nitrogen concentration profiles, reported different diffusion coefficients, indicating more than one diffusion route [12]. Moreover, they show that interstitial nitrogen diffuses much faster than substitutional nitrogen. In this case, the quantitative concentration of nitrogen obtained by EELS should be compared to the interstitial nitrogen concentration, which, from our *in situ* XPS results, is about 3at.%, meaning a better agreement between EELS nitrogen concentration near the surface and XPS results. It must be noted that since only about half to one third of the film is actually doped, the average nitrogen concentration is probably closer to 1-2at.%.

## 3.2 Electrical characterization

Very generally, the effect of nitrogen implantation and diffusion into the Anatase thin film can be monitored by 4 probe electrical resistivity measurements. Such results, converted into sheet resistance, are shown in Figure 8 (a). It is observed that sheet resistance drops by 7 orders of magnitude and reaches 54.7kΩ/□. These results are in close agreement with resistivity, ρ, as measured by van der Pauw method using indium contacts, shown in Figure 8(b). The films resistivity could be as low as $3·10^{-1}$ Ωcm, for the sample prepared at 500°C and implanted for 60 minutes (average nitrogen concentration about 1-2%). This resistivity is about 10 fold lower than reported for $TiO_{1.88}N_{0.12}$ (4at.% of nitrogen) prepared by plasma-assisted molecular beam epitaxy [33]. Moreover, the presented results are very similar to $N:TiO_2$ prepared by electron cyclotron resonance and by reactive sputtering in refs [4,6] (with slightly lower light transmittance, see below) but higher than $TiO_2$ doped with Nb, Ta or W, which may show resistivity much lower than $10^{-2}$ Ωcm [5,34,35]. In Figures 8 and 9 the symbols cover the estimated uncertainty bars.

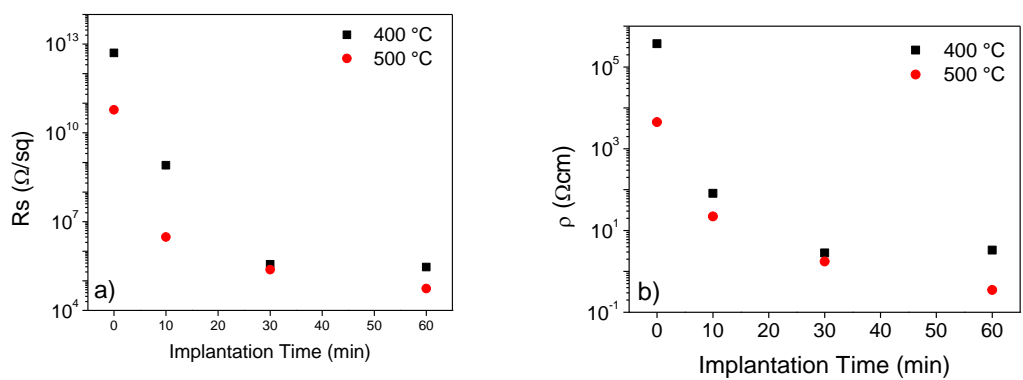

Figure 8: (a) Sheet resistance and (b) Resistivity from all samples measured by 4-probe and van der Pauw, respectively. The results are in close agreement.

Mobility and carrier concentration results are shown in Figure 9. Highest carrier concentration is observed for sample implanted at 500°C and for 60 minutes and reaches up to $6·10^{19}$ cm$^{-3}$. Mobility values measured are always lower than 1 $cm^2v^{-1}s^{-1}$ (and just above 0.1 $cm^2v^{-1}s^{-1}$), which is much lower (at least by one or even two orders of magnitude) than usual TCOs like ITO or $SnO_2$ [36]. Such mobility is however similar to reported to Nb doped $TiO_2$ [34]. Note that resistivity and carrier concentration values are calculated considering the full film thickness and, as EELS Nitrogen profile showed (Figure 6(a)), nitrogen concentration is far from homogenous along the thin film. If one takes into account that nitrogen is present in about one third of the films (30 nm instead of 90 nm), then carrier concentration would be in such region and in average about $2·10^{20}$ cm$^{-3}$ (it should be higher close to the surface). Such corrected carrier concentration starts to be similar to values obtained in the literature as $3·10^{20}$cm$^{-3}$ for $Ta:TiO_2$ and $10^{21}$ cm$^{-3}$ for $Nb:TiO_2$ [35,34]. Industry standard TCEs have again similar carrier concentration values, such as $1.5 \cdot 10^{20}$ cm$^{-3}$ for FTO, or about $20^{21}$ cm$^{-3}$ for ITO and AZO [37, 38, 39]. Similarly, sheet resistance (or resistivity) in the doped region would be 3 fold smaller than in the thin film average. Moreover, nitrogen concentration gradient may explain low mobility since the region more relevant to electrical measurements has higher carrier concentration, which in turn may reduce mobility.

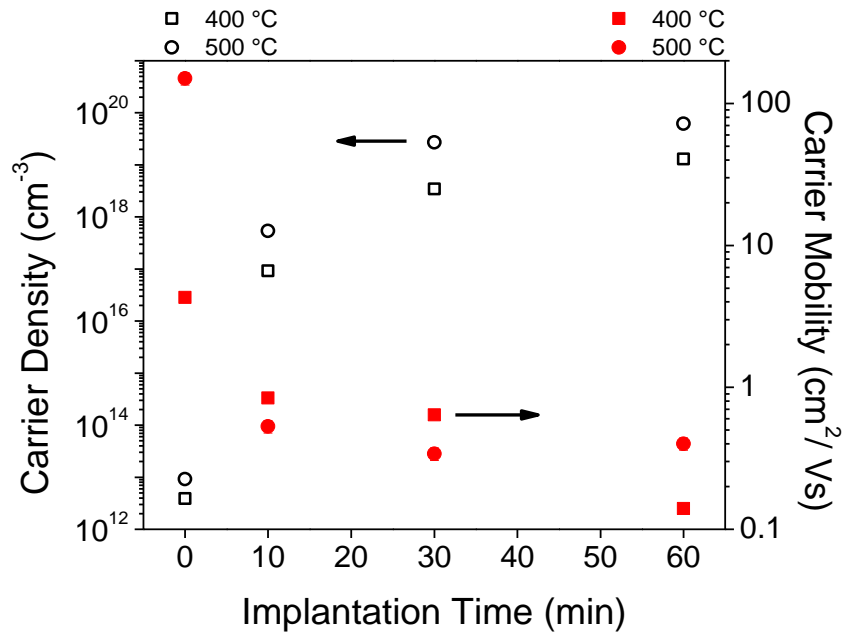

Figure 9: Hall mobility and carrier concentration measured by van der Pauw method. Carrier density is show in open symbols while solid symbols show carrier mobility.

The temperature dependence of the resistance is shown in Figure 10 (a) for the sample prepared at 500°C and implanted for 60 minutes. A very small hysteresis was observed. Failure to fit data in an Arrhenius plot (LnR vs. 1/T) denoted that the resistance is not governed by thermal activation, contrary to plasma-assisted molecular beam epitaxy N:TiO$_2$ samples that contained mostly substitutional nitrogen [33]. However, the resistance scales as LogR α T$^{-1/2}$, as shown in Figure 10 (b). This suggests that the conduction mechanism close to room temperature is variable range hopping (VRH). For this regime the resistivity should follow:

$$\rho(T) = \rho_0 \exp[T_0/T]^p, \qquad (1)$$

where p = 1/4 for Mott (Mott-VRH) [40] and p = 1/2 for Efros and Shklovskii (ES-VRH) [41]. Both mechanisms were observed in ion implanted TiO$_2$ single crystals [42] and disordered TiO$_2$ thin films [43,44] in a wide temperature range. To determine which one of the mechanisms is dominant in our sample we used the method proposed by Zabrodskii and Zinoveva [45] to obtain the exponent p, where w(T) = - ∂log(R)/∂log(T) and log(w) = log(pT0 p) – p log(T). By plotting log(w) versus log(T) we can find the value of the exponent p from the slope of the curve. As depicted in the insert of Figure 10(b), for T > 235 K the curve is fitted with p = 0.488 ± 0.007, very close to value expected for ES-VRH conduction mechanism. For lower temperatures, the data diverge, indicating a change in the conduction mechanism. Further study is necessary to understand this behavior but is outside the scope of this work.

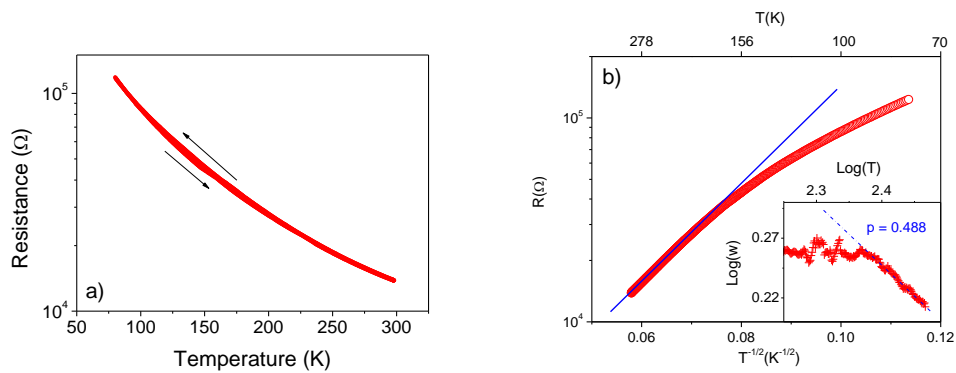

Figure 10. (a) Resistance versus temperature for the film prepared at 500°C and implanted for 60 minutes. A small hysteresis was observed. (b) Logarithm plot of resistance versus $T^{-1/2}$ showing a linear dependency (blue line) at high temperature. The inset shows the double-log plot of function w(T) and T and the value of exponent "p", as determined from the slope of the curve.

## 3.3 Optical Characterization

The UV-Vis-NIR light transmission spectra for the studied samples are shown in Figure 11. The transmission spectra for amorphous quartz (used as substrate) and undoped anatase prepared at 400 and 500° are also shown. Interference fringes are observed for the thin films and the maximum transmission is in the range from 500 to 600 nm (green). It is observed that undoped anatase have a transmission maximum very similar to amorphous quartz and that by doping the thin films the transmission falls from about 90% to 85% (with respect to air). Such observed transmission is better than some literature results for doped $TiO_2$ with similar resistivity, as indicated above. [4,6,34]. The transmission curves were simulated (not shown) using the method described in ref [46] and the general shape is very well described considering only the thickness and refraction indexes of the film and substrate. Transmission spectra measured further into the IR up to 3000 nm (not shown) are still featureless with only one absorption region near 2720 nm due to the quartz substrate.

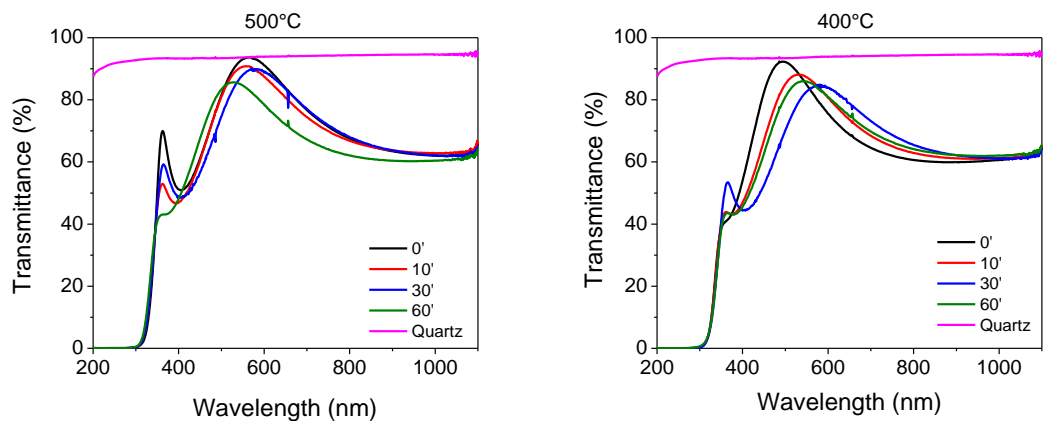

Figure 11: Light transmission for samples prepared at 500 and 400°C. Amorphous quartz substrate and undoped anatase thin film are also shown.

Absorption spectra can be used to determine the optical band-gap. Taking in account that Anatase has an indirect band-gap and following the procedure indicated in [47], the optical band gap can be obtained by plotting the square-root of the absorption coefficient as function of the energy and extrapolating the absorption edge at high energies [48]. Figure 12 shows the results for two extreme cases: undoped Anatase thin film prepared at 500°C and nitrogen implanted for 60 minutes also prepared at 500°C. Optical band-gap in both cases is about 3.29 eV, in agreement with Anatase value [47,49]. This indicates that the doped region does not affect significantly the overall light absorption edge. Similar results were obtained for all other samples (not shown). Such result is in agreement to literature reports that indicate that gap narrowing is related to substitutional nitrogen, which in our case could be restricted to the surface. Interstitial nitrogen, on the other hand, does not reduce band-gap [12,50,51].

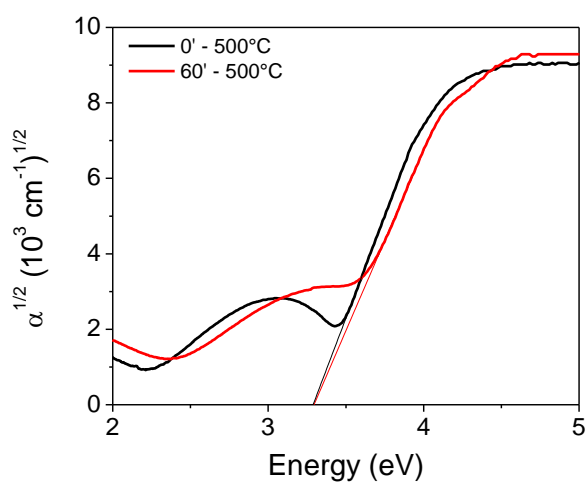

Figure 12: Square-root of the absorption coefficient for undoped and 60 minutes N doped Anatase films prepared at 500°C. In both cases, the gap is about 3.29 eV.

### 3.4 Shelf stability

Finally, the shelf stability was evaluated by measuring the resistivity, mobility and carrier concentration on the interval of some days for the sample implanted for 30 minutes at 500°C (without any surface protection/coating). The results, shown in Figure 13, indicate that the films maintains its resistivity with only a 40% resistivity increase, despite the fact that $TiO_2$ is thermodynamically favorable with respect to TiN. It is observed that the resistivity increases slightly from (1.8±0.2) to (2.5±0.3) $\Omega$cm, see Figure 13 (a). This change is accompanied by a decrease of carrier density and an increase in mobility, as shown in Figure 13(b) and (c). Such changes could be due to surface oxidation that would displace nitrogen and reduce its concentration. However, oxygen diffusion would be too slow to displace nitrogen deeper in the thin film. The stability of this $N:TiO_2$ film, even if not subjected to heat or UV light, is interesting with respect to literature [52]. Further study is needed to compare the stability of $N:TiO_2$ to that of other TCOs [53].

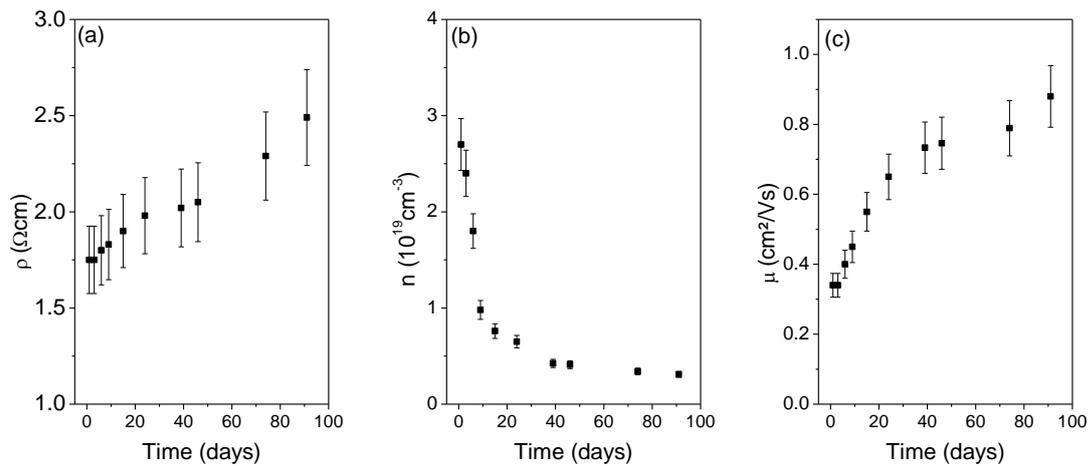

Figure 13: Resistivity, carrier concentration and mobility for shelf storage of the sample nitride for 30 minutes and prepared at 500°C.

# 4 Conclusions

In summary, a comprehensive study showed that, by implanting 150 eV nitrogen and hydrogen ions into Anatase films, it is possible to build ~33% nitrogen surface concentrations which drive nitrogen diffusion into the volume of the film. For the sample prepared at 500°C and implanted for 60 minutes, the films resistivity could be as low as $3 \cdot 10^{-1}$ Ωcm while transparency at 550nm is about 85%. In this case, nitrogen diffusion could reach about 25 to 30 nm deep into the thin film. Note that 150eV nitrogen ions are readily available with simple laboratory ion guns.

The proposed two step deposition and doping technique could, as planned, provide Anatase thin films with a nitrogen doped zone near the surface. Moreover, carrier densities and conductivities similar to other established TCOs could be obtained. These results show the effectiveness of nitrogen diffusion into the Anatase film from the surface due to the obtained nitrogen surface concentration and applied temperature. Clearly, it was not possible to dope with nitrogen the whole film or to create homogenously doped sample at 500°C and 60 minutes of implantation. However, by adjusting properly the nitriding implanting time, desired thin film properties could be obtained. Indeed, the presented results indicate that, by doping during longer times until the whole thin film is doped, it could be possible to obtain resistivities lower than $10^{-1}$ Ωcm.

This study supports the interpretations that interstitial nitrogen has higher binding energy in XPS, that it diffuses faster in Anatase (with respect to Nitrogen in substitutional sites) and that it does not affect Anatase optical bandgap.

Finally, low energy ion doping using simple ion guns can be applied to Anatase prepared by other means, even colloidal synthesized nanoparticles. The high reactivity of low energy nitrogen ions associated with hydrogen ions, we speculate, could be also effective in other Anatase samples kinds.

# 5 Acknowledgments

Part of this work was supported by FAPESP, projects 2014/23399-9 and 2012/10127-5. TEM experiments were performed at the Brazilian Nanotechnology National Laboratory (LNNano/CNPEM).